\documentclass[aps,prl,twocolumn,showpacs]{revtex4}
\newcommand{\ra}{\rangle}
\newcommand{\la}{\langle}

\usepackage{graphicx}
\begin{document}

\title{\bf Generation of Bright Two-Color Continuous Variable 
Entanglement}
\author{A. S. Villar, L. S. Cruz, K. N. Cassemiro, M. Martinelli, and P.
Nussenzveig}\email[E-mail: ]{nussen@if.usp.br}
\affiliation{Instituto de F\'\i sica, Universidade de S\~ao Paulo,
Caixa Postal 66318, 05315-970 S\~ao Paulo, SP, Brazil}
\date{\today}

\begin{abstract}
We present the first measurement of squeezed-state entanglement 
between the twin beams produced in an Optical Parametric Oscillator 
(OPO) operating above threshold. Besides the usual squeezing in the 
intensity difference between the twin beams, we have measured squeezing in 
the sum of phase quadratures. Our scheme enables us to 
measure such phase anti-correlations between fields of different 
frequencies. In the present measurements, wavelengths differ 
by $\approx 1$~nm.  Entanglement is demonstrated according to the 
Duan {\em et al.} criterion~[Phys. Rev. Lett. {\bf 84}, 2722 
(2000)] $\Delta^2\hat{p}_- +\Delta^2\hat{q}_+=1.47(2)<2$. This 
experiment opens the way for new potential applications such as the 
transfer of quantum information between different parts of the 
electromagnetic spectrum.

\end{abstract}

\maketitle

The field of quantum information has recently attracted great 
interest, owing to potential applications in information storage, 
communication, and computing~\cite{nielsenchuang}. Entanglement is 
viewed as a key resource for these applications, especially for quantum 
communication. A variety of physical systems presenting entanglement 
have been investigated both theoretically and experimentally. The 
vast majority of experiments concentrate on discrete variable systems, 
such as trapped ions~\cite{trappedions}, few photon electromagnetic fields 
in cavity QED~\cite{cavityQED}, spontaneous parametric 
downconversion~\cite{downconversion}, and nuclear magnetic 
resonance~\cite{NMR}. On the other hand, in recent experiments, 
continuous variable systems have been studied, including light 
beams~\cite{kimbleepr92} and atomic samples~\cite{CVatomic}. The 
number of such experiments, however, is still relatively small in 
comparison with discrete variable systems.

The first experimental demonstration of continuous variable entanglement 
used a continuous-wave (CW) optical parametric oscillator (OPO) operating 
below threshold~\cite{kimbleepr92}. The two squeezed vacuum outputs were 
shown to possess Einstein Podolsky Rosen (EPR) type correlations. Most 
recent experiments use a nonlinear interaction to produce squeezed fields, 
which are then combined in a beamsplitter to generate 
entanglement~\cite{entangBS}. Conversely, a beamsplitter transformation 
can also be used to generate a squeezed beam from an entangled 
input~\cite{pengpra00}. In these experiments, it is mandatory to 
have fields of the same frequency. 

In this Letter, we present the first (to our knowledge) measurement 
of continuous variable entanglement between bright fields of truly different 
frequencies. Even before the first experiment~\cite{kimbleepr92}, it 
was predicted that the above-threshold OPO should produce entangled twin 
beams~\cite{reiddrummprl88,reiddrummpra89}. So far, this prediction had 
not yet been verified, owing to the difficulty of measuring phase 
properties of the twin beams. We have succeeded in measuring quantum 
correlations between the sum of phase quadratures of non-degenerate 
twin beams. 

Bipartite continuous variable entanglement~\cite{braunsteinreview} 
can be tested according to a criterion established by Duan {\it et 
al.}~\cite{dgcz} and also by Simon~\cite{simon}. The criterion 
is based on the total variance of EPR-type 
operators. For operators $\hat{x}_i$ and $\hat{p}_i$ that obey 
position-momentum commutation relations, they consider the variances 
of combinations such as $\hat{x}_1 + \hat{x}_2$ and 
$\hat{p}_1 - \hat{p}_2$. The quadratures of electromagnetic fields 
satisfy such commutation relations. We focus here on the 
so-called amplitude $\hat{p}$ and phase $\hat{q}$ field quadratures, 
defined by: 
\begin{equation} 
\hat{a} = \frac{e^{i\phi}}{2} (\hat{p}+i\hat{q}) \;, 
\label{eq:quadr_def} 
\end{equation} 
where $\hat{a}$ is the field annihilation operator, $\phi$ is an 
arbitrary phase and, for a macroscopic field with a well-defined 
mean amplitude, $\la\hat{q}\ra = 0$. In terms of these operators, 
inseparability (entanglement) is demonstrated by a violation of 
the inequality: 
\begin{equation} 
\Delta^2 \left(\frac{\hat{p}_1 - \hat{p}_2}{\sqrt{2}}\right) + \Delta^2 
\left(\frac{\hat{q}_1 + \hat{q}_2}{\sqrt{2}}\right) \ge 2 \;, 
\label{eq:dgcz}
\end{equation} 
where the Standard Quantum Level (SQL) is normalized to 1 for 
each combination of quadratures. In order to simplify notation, we 
will refer to $(\hat{p}_1 \pm \hat{p}_2)/\sqrt{2}$ as $\hat{p}_{\pm}$, 
and to $(\hat{q}_1 \pm \hat{q}_2)/\sqrt{2}$ as $\hat{q}_{\pm}$ [thus, 
$\Delta^2 \hat{p}_-<1$ signals squeezing in the difference of amplitude 
quadratures]. If both $\hat{p}_-$ and $\hat{q}_+$ are squeezed, 
inequality (\ref{eq:dgcz}) is violated and we have squeezed-state 
entanglement.

It is easy to understand why entanglement is expected in the CW OPO 
operating above threshold. The OPO consists of a $\chi^{(2)}$ nonlinear 
crystal inside a resonant cavity, in which parametric downconversion 
takes place. The cavity feeds back the downconverted fields, leading 
to stimulated parametric gain and hence to an oscillation threshold. 
Since the primary downconversion process involves creating pairs of 
photons by annihilation of pump photons, one naturally expects strong 
correlations between the intensities of the twin beams: positive 
intensity fluctuations of one beam correspond to positive intensity 
fluctuations of the other beam. These correlations, however, are 
frequency-dependent. For power spectrum analysis frequencies larger 
than the OPO cavity bandwidth, correlations tend to disappear since, 
for times shorter than the cavity lifetime, a photon can exit the 
cavity while its ``twin'' still remains inside. Squeezing in the 
intensity difference was already observed back in 1987~\cite{twinopo87}. 
On the other hand, energy conservation ($\omega_0 = \omega_1 + 
\omega_2$, where indices 0, 1, and 2 refer to pump, signal and 
idler beams, respectively) and phase matching imply strong 
anti-correlations between phase fluctuations: positive phase 
fluctuations of one beam correspond to negative phase fluctuations 
of its ``twin''. This is exactly the situation discussed 
following~(\ref{eq:dgcz}): entanglement occurs if these 
(anti)correlations lead to fluctuations below the SQL. However, 
squeezing in the phase-sum had not been measured to date.

The first prediction of entanglement in the above-threshold OPO was 
made by Reid and Drummond~\cite{reiddrummprl88}, who even 
suggested a way of measuring phase fluctuations and analyzed the low 
frequency effects of phase diffusion~\cite{reiddrummpra89}. A recent 
detailed prediction, taking into account the effects of pump noise and 
cavity detunings for the three fields, for a triply resonant 
OPO, was presented in~\cite{optcomm04}. The basic difficulty in measuring 
(phase) quadrature fluctuations, in contrast with intensity fluctuations, 
is that the standard technique, homodyne detection, calls 
for a local oscillator having a well-defined phase relationship 
with the field to be measured. In the OPO, this is difficult to 
implement, since the frequencies of the twin beams are usually 
different, depending on the oscillating modes, and vary from one 
realization to the next. Hence, two local oscillators would be 
required: one for each beam. It would also be necessary to 
phase-lock these fields to the twin beams.  

One way to overcome this difficulty is to force the OPO 
to oscillate in a strictly frequency-degenerate situation. This is 
technically challenging and has been done by two groups, using different 
approaches~\cite{pfisteropodegen, claudelamina}. Our strategy is to 
perform self-homodyne measurements, without the use of local oscillators, by 
a frequency-dependent reflection of each beam~\cite{levenson}. If one 
considers the field as a mean value at a carrier frequency with noise 
sidebands at some analysis frequency, a frequency-dependent reflectivity 
entails different phase shifts for the carrier and sidebands. Consequently, 
different quadrature fluctuations can be projected onto amplitude 
fluctuations (with respect to the phase of the mean field). A detailed 
description for a single field reflected off an optical cavity was 
given by Galatola {\it et al.}~\cite{galatola}. For an imperfect cavity, 
in which vacuum leaks from the outside through the mirrors, the 
reflected beam amplitude noise power spectrum $S_R(\Omega)$ can be 
written as~\cite{optcomm04}:
\begin{equation}
S_R(\Omega)=|g_p|^2\,S_p(\Omega)+|g_q|^2\,S_q(\Omega)+|g_{vp}|^2+|g_{vq}|^2\;,
\label{eq:S_R} 
\end{equation}
where $S_p(\Omega)$ and $S_q(\Omega)$ are the incident beam amplitude 
and phase noise, respectively, and $g_p$, $g_q$, $g_{vp}$ and $g_{vq}$ 
are coefficients that depend on cavity reflection and transmission 
coefficients through the relations:
\begin{eqnarray}
\label{eq:coef_g} 
g_p&=&\frac{1}{2}\left[\frac{r^*(\Delta)}{|r(\Delta)|}\,r(\Delta+\Omega)
+\frac{r(\Delta)}{|r(\Delta)|}\,r^*(\Delta-\Omega)\right]\\ \nonumber
g_q&=&\frac{1}{2}\left[\frac{r^*(\Delta)}{|r(\Delta)|}\,r(\Delta+\Omega)
-\frac{r(\Delta)}{|r(\Delta)|}\,r^*(\Delta-\Omega)\right]\\ \nonumber
g_{vp}&=&\frac{1}{2}\left[\frac{t^*(\Delta)}{|t(\Delta)|}\,t(\Delta
+\Omega)+\frac{t(\Delta)}{|t(\Delta)|}\,t^*(\Delta-\Omega)\right]\\ \nonumber
g_{vq}&=&\frac{1}{2}\left[\frac{t^*(\Delta)}{|t(\Delta)|}\,t(\Delta+
\Omega)-\frac{t(\Delta)}{|t(\Delta)|}\,t^*(\Delta-\Omega)\right]\;.
\end{eqnarray}
Amplitude reflection $r(\Delta)$ and transmission $t(\Delta)$ coefficients 
can be simply written as:
\begin{eqnarray}
\label{eq:coef_rt} 
r(\Delta)&=&\frac{r_1-r_2\,\exp(i\Delta/\delta\nu_{ac})}{1-r_1\,r_2\,
\exp(i\Delta/\delta\nu_{ac})}\;,\\ \nonumber
t(\Delta)&=&\frac{t_1\,t_2\,\exp(i\Delta/\delta\nu_{ac})}{1-r_1\,r_2\,
\exp(i\Delta/\delta\nu_{ac})}\;,
\end{eqnarray}
where $\Delta$ is the detuning between the incident field central 
frequency and the cavity resonance frequency, 
$\delta\nu_{\mbox{\scriptsize ac}}$ is the cavity bandwidth (FWHM), 
and $\Omega$ is the analysis frequency. Cavity input mirror amplitude 
reflection and transmission coefficients are denoted 
by $r_1$ and $t_1$, while $t_2$ is defined so that all internal 
losses $A$ obey the relation $t_2^2=A=1-r_2^2$. 

For analysis frequencies larger than $\sqrt{2}\,\delta
\nu_{\mbox{\scriptsize ac}}$, it is possible to completely convert 
incident phase fluctuations into amplitude fluctuations of the 
reflected beam. As a matter of fact, if this condition is satisfied, 
then, for $\Delta=0$, $|g_{p}|^2\approx 1$ and we recover amplitude 
fluctuations. Amplitude fluctuations are also recovered for $\Delta 
\gg \delta\nu_{\mbox{\scriptsize ac}}$, for which $|g_{p}|^2\rightarrow 
1$. For $\Delta = \pm\delta \nu_{\mbox{\scriptsize ac}}/2$, then 
$g_{q}\approx 1$ and incident phase fluctuations are projected onto 
amplitude fluctuations of the reflected beam. We use one analysis 
cavity for each beam and scan their frequencies synchronously. In 
this way, we are always measuring the same quadrature for each field, 
with respect to its mean value, {\it regardless of the frequency 
difference between the fields}. 

\begin{figure}[h]
\includegraphics[width=8.5cm]{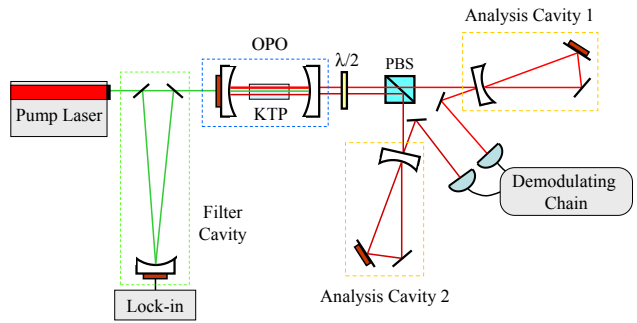}
\caption{Sketch of the experimental setup (details explained in the text).}
\label{fig:expset}
\end{figure}

Our experiment is performed with a triply resonant CW OPO operating above 
threshold~\cite{bjp01}. The pump laser is an ultrastable diode-pumped 
frequency-doubled Nd:YAG source (Innolight Diabolo) at 532~nm, with a 
second output beam at 1064~nm, which we use for alignment purposes. 
The nonlinear crystal is a 12~mm-long type-II High Gray Tracking Resistant KTP 
from Raicol. Crystal temperature is kept near 24$^{\circ}$C, with a 
stability of the order of 10~mK, by means of a peltier element. The 
cavity is a quasi-confocal linear Fabry Perot cavity, with input mirror 
reflectivities equal to 89\% at 532~nm and greater than 99.8\% at 
1064~nm. Output mirror reflectivities are greater than 99.8\% at 
532~nm and 95\% at 1064~nm. The typical threshold power is 60~mW and OPO 
cavity bandwidth is $\delta\nu_{\mbox{\tiny OPO}} = 52(1)$~MHz. 
Noise in the difference of signal and idler intensities 
has a stable value $\Delta^2 \hat{p}_-=0.63(1)$ registered by the 
photodetectors, or $-2.0(7)$~dB. According to~\cite{optcomm04}, 
$\Delta^2 \hat{q}_+$ is significantly affected by pump excess phase 
noise. Our laser presented excess noise up to 25~MHz, which we 
had to filter by transmission through a ring cavity, with a bandwith 
$\delta\nu_{\mbox{\scriptsize f}} = 2.3(1)$~MHz. In this way, the pump 
beam was shot-noise-limited for frequencies above 15~MHz. 

The experimental setup is sketched in fig.~\ref{fig:expset}. The pump 
beam at 532~nm is sent through the filter cavity and then mode-matched 
to the OPO cavity. The orthogonally polarized twin beams produced, with 
powers of the order of a few milliwatts each, are separated by a polarizing 
beamsplitter cube (PBS) and each directed to a tunable ring analysis cavity. 
For our working crystal temperature, wavelengths of signal and idler 
beams can differ by 0.8~nm to 0.9~nm. Analysis cavity bandwidths are 
$\delta\nu_{\mbox{\scriptsize ac}} = 14(1)$~MHz. The reflected field 
is detected by a high quantum efficiency [85(3)\%] photodiode (Epitaxx 
ETX 100). The photocurrent is preamplified and the DC and High Frequency (HF) 
components separated. The HF components are sent to a demodulating chain, 
where they are mixed with a sinusoidal reference at the analysis frequency 
$\Omega=27$~MHz (RBW of 600~kHz). As the analysis cavities' resonances are 
swept over time, variances of each individual noise component, of their sum, 
and of their difference are calculated. The number $N$ of points used, which is 
proportional to the acquisition time, is large enough to guarantee a well 
defined variance and small enough to correspond to a very small change in 
cavities' detuning (this is equivalent to a spectrum analyzer VBW, in our 
case approximately 1~kHz).

\begin{figure}[th]
\includegraphics[width=8.5cm]{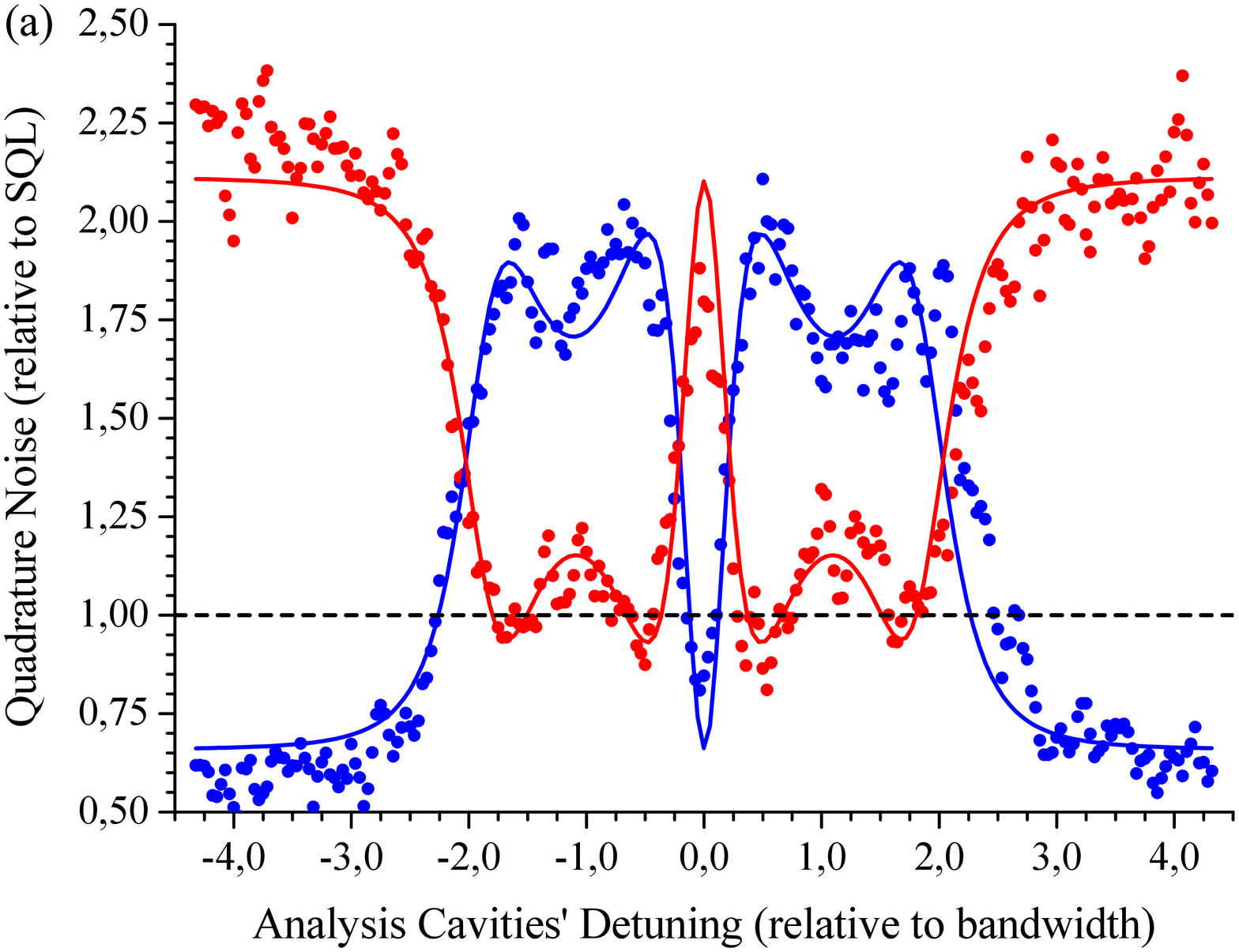}
\includegraphics[width=8.5cm]{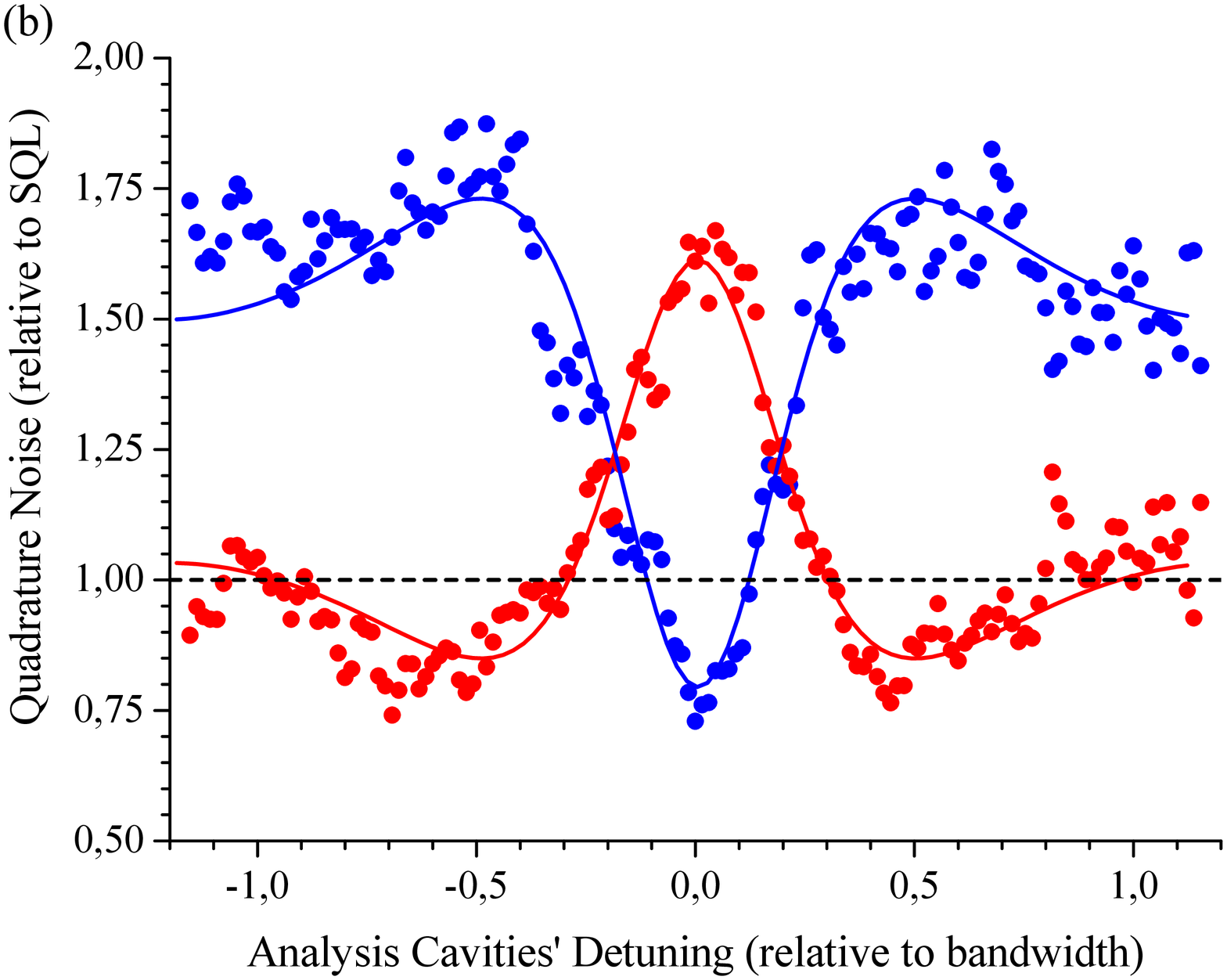}
\caption{Measurements of sum (red) and difference (blue) of quadrature 
fluctuations as functions of analysis cavities' detunings. In (a) the detuning 
spans the range $\pm 4 \; \delta\nu_{\mbox{\scriptsize ac}}$. In (b), a local 
scan is presented, in which we clearly observe $\Delta^2\hat{p}_- < 1$ 
for $\Delta=0$ and $\Delta^2\hat{q}_+ < 1$ for $|\Delta| = 0.5 \; \delta 
\nu_{\mbox{\scriptsize ac}}$, characterizing squeezed-state 
entanglement. The solid curves are theoretical fits to the data (eq.\ref{eq:S_R}).}
\label{fig:expresults}
\end{figure}

Sum and difference noise spectra recorded as functions of the synchronous 
analysis cavity frequency scans are presented in fig.~\ref{fig:expresults}. 
A scan over $\pm 4 \; \delta\nu_{\mbox{\scriptsize ac}}$ is presented 
in fig.~\ref{fig:expresults}~(a). From the sum and difference spectra, one 
can easily notice that the quadrature being measured on each beam alternates 
between amplitude and phase. In particular, we recognize the amplitude 
quadrature behavior for detunings $|\Delta| \ge 3 \; \delta 
\nu_{\mbox{\scriptsize ac}}$ and for $\Delta=0$. One observes that 
$\Delta^2\hat{p}_-$ at $\Delta=0$ does not recover the squeezing observed 
for $|\Delta| \ge 3 \; \delta \nu_{\mbox{\scriptsize ac}}$, owing to 
a lack of experimental resolution (since the cavity frequency scan is 
large compared to $\delta\nu_{\mbox{\scriptsize ac}}$, very few data 
points are obtained at any particular detuning). Phase quadrature is 
measured for $|\Delta| = 0.5 \; \delta\nu_{\mbox{\scriptsize ac}}$ and, for 
the particular choice of analysis frequency we use, also at $|\Delta| = 
1.65 \; \delta\nu_{\mbox{\scriptsize ac}}$. For all other detunings we 
measure a linear combination of amplitude and phase quadratures. In 
fig.~\ref{fig:expresults}~(b), we present a local scan, with increased 
resolution, over $\pm 1 \; \delta\nu_{\mbox{\scriptsize ac}}$. Phase sum 
squeezing is observed, with $\Delta^2\hat{q}_+ = 0.84(2)$. 

The solid curves in fig.~\ref{fig:expresults}~(a) and (b) represent 
theoretical fits~\cite{optcomm04} to the data, taking into account the 
transfer of quadrature noise to amplitude noise of the fields reflected 
from the analysis cavities (eq.~\ref{eq:S_R}). Apart from scale factors 
and curve central position, the relevant free parameters 
are the combined quadrature fluctuations $\Delta^2\hat{p}_+$,
$\Delta^2\hat{p}_-$, $\Delta^2\hat{q}_+$, and $\Delta^2\hat{q}_-$. 
All other variables required in eq.~\ref{eq:S_R}, such as cavity bandwidth 
and analysis frequency, are independently measured and employed as 
constants in the fitting, which is done either for the sum 
or for the difference in each part of fig.~\ref{fig:expresults}, with 
equally good results.


In the first spectra we recorded, with pump power equivalent to 
1.5 times the threshold power, we observed strong excess noise in 
the phase sum, as was also found in~\cite{claudelamina}. This 
excess noise is not predicted by the standard OPO linearized 
theory~\cite{reiddrummprl88} and it is still not clear whether 
a full quantum theory~\cite{drummkaledpra02} can account for it. 
Although the calculations in~\cite{optcomm04} are carried out 
with the linearized theory, they provided us with useful 
indications: squeezing in the phase sum improves for nonzero 
detunings of signal and idler beams with respect to the 
OPO cavity. This behavior is also observed as pump power decreases, 
approaching the threshold power. Consistently with these indications, 
we could only observe phase sum squeezing very close to threshold 
(approximately seven percent above threshold). In this situation, 
the OPO is very unstable, hindering the measurements.

In terms of the entanglement criterion of eq.~\ref{eq:dgcz}, 
using $\Delta^2 \hat{q}_{+}=0.84(2)$, we obtain, 
$\Delta^2\hat{p}_- + \Delta^2\hat{q}_+ = 1.47(2)<2$, a 
clear violation of the inequality, characterizing, for the 
first time, {\it entanglement between bright beams of truly 
different frequencies}. Eq.~\ref{eq:dgcz} is a necessary 
and sufficient separability condition for gaussian states, which 
are predicted for the OPO and are consistent with our data. We have 
substantial losses from the output of the OPO to the detectors. 
If the overall detection efficiency is $\eta$, the measured 
variances are related to the ``true'' variances (e.g., 
$\Delta^2\hat{p}'$) by $\Delta^2\hat{p}=\eta(\Delta^2\hat{p}' 
-1)+1$. Correcting for losses, equal to 28(2)\%, we 
obtain $\Delta^2\hat{p}'_- + \Delta^2\hat{q}'_+ = 1.26(4)$. 
Another, more stringent criterion, is the so-called EPR 
criterion, which enables one to infer variances on one beam, 
as functions of variances on the other 
beam~\cite{reiddrummprl88}. This criterion implies 
a violation of the inequality $\Delta^2
\hat{p}_{\mbox{\scriptsize inf}} \; \Delta^2
\hat{q}_{\mbox{\scriptsize inf}} \ge 1$, where
\begin{equation}
\label{eq:varinf} 
\Delta^2\hat{p}_{\mbox{\scriptsize inf}}=\Delta^2\hat{p_1}
\left(1-\frac{\langle\delta\hat{p}_1\delta\hat{p}_2\rangle^2}{\Delta^2
\hat{p}_1\Delta^2\hat{p}_2}\right) \;, 
\end{equation}
with $\delta\hat{p}_i=\hat{p}_i-\la\hat{p}_i\ra$. An analogous relation holds 
for $\Delta^2\hat{q}_{\mbox{\scriptsize inf}}$. Our measured data does not 
violate the inequality, resulting in a product equal to $1.09(4)$. 
However, correcting for losses, the product is $\Delta^2
\hat{p}'_{\mbox{\scriptsize inf}} \; \Delta^2
\hat{q}'_{\mbox{\scriptsize inf}} = 0.91(5)$, violating this 
second criterion as well.

The entanglement generated in this system can be substantially 
improved. The intensity difference squeezing measured 
in~\cite{claudelamina} reached the impressive value of 9.7~dB, 
which would lead to even stronger violations of the above criteria. 
One other problem in our system is the unstable operation very 
close to threshold. However, very stable operation only a few percent 
above threshold can be obtained, as shown in~\cite{pfisteropodegen}. 

In summary, we have demonstrated, for the first time, bright 
two-color squeezed-state continuous variable entanglement. 
Applications to quantum information, such as quantum key 
distribution~\cite{qkdleuchsprl02} and quantum 
teleportation~\cite{kimbleteleport} with continuous variables, 
can be easily envisaged. Our measurement scheme has interesting 
properties for both. Quantum key distribution with squeezed-state 
entanglement usually requires sending the local oscillator, in addition 
to the entangled field, so as to enable homodyne measurements, 
a requirement which is not necessary in our case. Teleportation, 
on the other hand, has been restricted to fields of the same frequency. 
Two-color entanglement opens the way for distributing quantum 
information between different parts of the electromagnetic spectrum. 

\begin{acknowledgments} 
This work was funded by FAPESP, CAPES, and CNPq through 
{\it Instituto do Mil\^enio de Informa\c c\~ao Qu\^antica}. 
We gratefully acknowledge very useful and stimulating discussions 
with Claude Fabre.
\end{acknowledgments}


\end{document}